\documentclass[10pt, a4paper, twocolumn,backend=bibtex,style=authoryear,natbib=true,backend=biber]{article} 

\usepackage[english]{babel} 

\usepackage{microtype} 

\usepackage{amsmath,amsfonts,amsthm} 

\usepackage[svgnames]{xcolor} 

\usepackage[hang, small, labelfont=bf, up, textfont=it]{caption} 

\usepackage{booktabs} 

\usepackage{lastpage} 

\usepackage{graphicx} 

\usepackage{enumitem} 
\setlist{noitemsep} 

\usepackage{sectsty} 
\allsectionsfont{\usefont{OT1}{phv}{b}{n}} 

\usepackage{geometry} 

\usepackage[T1]{fontenc} 
\usepackage[utf8]{inputenc} 

\usepackage{XCharter} 

\usepackage{fancyhdr} 
\fancypagestyle{firstpage}{ 
	\fancyhf{}
}

\newcommand{\authorstyle}[1]{{\large\usefont{OT1}{phv}{b}{n}\color{DarkRed}#1}} 

\newcommand{\institution}[1]{{\footnotesize\usefont{OT1}{phv}{m}{sl}\color{Black}#1}} 

\usepackage{titling} 

\newcommand{\HorRule}{\color{DarkGoldenrod}\rule{\linewidth}{1pt}} 

\pretitle{
	\vspace{-30pt} 
	\HorRule\vspace{10pt} 
	\fontsize{32}{36}\usefont{OT1}{phv}{b}{n}\selectfont 
	\color{DarkRed} 
}

\posttitle{\par\vskip 15pt} 

\preauthor{} 

\postauthor{ 
	\vspace{10pt} 
	\par\HorRule 
	\vspace{20pt} 
}

\usepackage{lettrine} 
\usepackage{fix-cm}	

\newcommand{\initial}[1]{ 
	\lettrine[lines=3,findent=4pt,nindent=0pt]{
		\color{DarkGoldenrod}
		{#1}
	}{}%
}

\usepackage{xstring} 

\newcommand{\lettrineabstract}[1]{
	\StrLeft{#1}{1}[\firstletter] 
	\initial{\firstletter}\textbf{\StrGobbleLeft{#1}{1}} 
}

\usepackage{subcaption} 
\usepackage{hyperref}
\hypersetup{
    colorlinks=true,
    linkcolor=blue,
    filecolor=magenta,      
    urlcolor=cyan,
}
\usepackage[utf8]{inputenc}
\usepackage[english]{babel}

\usepackage[backend=bibtex,style=authoryear,natbib=true]{biblatex} 

\usepackage[autostyle=true]{csquotes} 

\def\aj{\ref@jnl{AJ}}                   
\def\actaa{\ref@jnl{Acta Astron.}}      
\def\araa{\ref@jnl{ARA\&A}}             
\def\apj{\ref@jnl{ApJ}}                 
\def\apjl{\ref@jnl{ApJ}}                
\def\apjs{\ref@jnl{ApJS}}               
\def\ao{\ref@jnl{Appl.~Opt.}}           
\def\apss{\ref@jnl{Ap\&SS}}             
\def\aap{\ref@jnl{A\&A}}                
\def\aapr{\ref@jnl{A\&A~Rev.}}          
\def\aaps{\ref@jnl{A\&AS}}              
\def\azh{\ref@jnl{AZh}}                 
\def\baas{\ref@jnl{BAAS}}               
\def\bac{\ref@jnl{Bull. astr. Inst. Czechosl.}}
\def\caa{\ref@jnl{Chinese Astron. Astrophys.}}
\def\cjaa{\ref@jnl{Chinese J. Astron. Astrophys.}}
\def\icarus{\ref@jnl{Icarus}}           
\def\jcap{\ref@jnl{J. Cosmology Astropart. Phys.}}
\def\jrasc{\ref@jnl{JRASC}}             
\def\memras{\ref@jnl{MmRAS}}            
\def\mnras{\ref@jnl{MNRAS}}             
\def\na{\ref@jnl{New A}}                
\def\nar{\ref@jnl{New A Rev.}}          
\def\pra{\ref@jnl{Phys.~Rev.~A}}        
\def\prb{\ref@jnl{Phys.~Rev.~B}}        
\def\prc{\ref@jnl{Phys.~Rev.~C}}        
\def\prd{\ref@jnl{Phys.~Rev.~D}}        
\def\pre{\ref@jnl{Phys.~Rev.~E}}        
\def\prl{\ref@jnl{Phys.~Rev.~Lett.}}    
\def\pasa{\ref@jnl{PASA}}               
\def\pasp{\ref@jnl{PASP}}               
\def\pasj{\ref@jnl{PASJ}}               
\def\rmxaa{\ref@jnl{Rev. Mexicana Astron. Astrofis.}}%
\def\qjras{\ref@jnl{QJRAS}}             
\def\skytel{\ref@jnl{S\&T}}             
\def\solphys{\ref@jnl{Sol.~Phys.}}      
\def\sovast{\ref@jnl{Soviet~Ast.}}      
\def\ssr{\ref@jnl{Space~Sci.~Rev.}}     
\def\zap{\ref@jnl{ZAp}}                 
\def\nat{\ref@jnl{Nature}}              
\def\iaucirc{\ref@jnl{IAU~Circ.}}       
\def\aplett{\ref@jnl{Astrophys.~Lett.}} 
\def\apspr{\ref@jnl{Astrophys.~Space~Phys.~Res.}}
\def\bain{\ref@jnl{Bull.~Astron.~Inst.~Netherlands}} 
\def\fcp{\ref@jnl{Fund.~Cosmic~Phys.}}  
\def\gca{\ref@jnl{Geochim.~Cosmochim.~Acta}}   
\def\grl{\ref@jnl{Geophys.~Res.~Lett.}} 
\def\jcp{\ref@jnl{J.~Chem.~Phys.}}      
\def\jgr{\ref@jnl{J.~Geophys.~Res.}}    
\def\jqsrt{\ref@jnl{J.~Quant.~Spec.~Radiat.~Transf.}}
\def\memsai{\ref@jnl{Mem.~Soc.~Astron.~Italiana}}
\def\nphysa{\ref@jnl{Nucl.~Phys.~A}}   
\def\physrep{\ref@jnl{Phys.~Rep.}}   
\def\physscr{\ref@jnl{Phys.~Scr}}   
\def\planss{\ref@jnl{Planet.~Space~Sci.}}   
\def\procspie{\ref@jnl{Proc.~SPIE}}   

\title{The TRAPPIST-1\\JWST Community Initiative} 

\author{
	\authorstyle{Micha\"el Gillon\textsuperscript{1},
	Victoria  Meadows\textsuperscript{2}, 
	Eric Agol\textsuperscript{2}, 
	Adam J.\ Burgasser\textsuperscript{3}, 
	Drake Deming\textsuperscript{4}, 
	Ren\'e Doyon\textsuperscript{5},
	Jonathan Fortney\textsuperscript{6},
	Laura Kreidberg\textsuperscript{7}, 
	James Owen\textsuperscript{8}, 
	Franck Selsis\textsuperscript{9}, 
	Julien de Wit\textsuperscript{10},
	Jacob Lustig-Yaeger\textsuperscript{2},
	Benjamin V.\ Rackham\textsuperscript{10}
	} 
	\newline\newline 
	\textsuperscript{1}\institution{Astrobiology Research Unit, University of Li\`ege, Belgium}\\ 
	\textsuperscript{2}\institution{Department of Astronomy, University of Washington, USA}\\ 
	\textsuperscript{3}\institution{Department of Physics, University of California San Diego, USA}\\ 
	\textsuperscript{4}\institution{Department of Astronomy, University of Maryland at College Park, USA}\\ 
    \textsuperscript{5}\institution{Institute for Research in Exoplanets, University of Montreal, Canada}\\ 
    \textsuperscript{6}\institution{Other Worlds Laboratory, University of California Santa Cruz, USA}\\ 
    \textsuperscript{7}\institution{Center for Astrophysics | Harvard and Smithsonian,  USA}\\ 
    \textsuperscript{8}\institution{Department of Physics, Imperial College London, United Kingdom}\\ 
    \textsuperscript{9}\institution{Laboratoire d'Astrophysique de Bordeaux, University of Bordeaux, France} \\ 
    \textsuperscript{10}\institution{Department of Earth, Atmospheric, and Planetary Sciences, MIT, USA} 
}

\date{\today} 

\begin{document}

\maketitle 

\thispagestyle{firstpage} 


\lettrineabstract{The upcoming launch of the  James Webb Space Telescope (JWST) combined with the unique features of the TRAPPIST-1 planetary system should enable the young field of exoplanetology to enter into the realm of temperate Earth-sized worlds. Indeed, the proximity of the system (12pc) and the small size (0.12 $R_\odot$) and luminosity (0.05\% $L_\odot$) of its host star should make the comparative atmospheric characterization of its seven transiting planets within reach of an ambitious JWST program. Given the limited lifetime of JWST, the ecliptic location of the star that limits its visibility to 100d per year, the large number of observational time required by this study, and the numerous observational and theoretical challenges awaiting it, its full success will critically depend on a large level of coordination between the involved teams and on the support of a large community. In this context, we present here a community initiative aiming to develop a well-defined sequential structure for the study of the system with JWST and to coordinate on every aspect of its preparation and implementation, both on the observational (e.g. study of the instrumental limitations, data analysis techniques, complementary space-based and ground-based observations) and theoretical levels (e.g. model developments and comparison, retrieval techniques, inferences).  Depending on the outcome of the first phase of JWST observations of the planets, this initiative could become the seed of a major JWST Legacy Program devoted to the study of TRAPPIST-1.}
\\
\\
\noindent \textbf{Keywords:} planetary systems, star and planet formation, stars and stellar evolution 

\section{JWST \& TRAPPIST-1: towards the spectroscopic study of potentially habitable exoplanets}

Since the seminal discovery of 51 Pegasi b \citep{Mayor1995},  more than four  thousands exoplanets have been detected at an ever increasing rate, including a steeply growing fraction of planets significantly smaller than Uranus and Neptune, some even smaller than our Earth. In parallel, many projects aiming   to {\it characterize}  exoplanets have developed during the last two decades, bringing notably first pieces of information on the atmospheric properties of several dozens of exoplanets. Most of these atmospheric studies have been made possible by the transiting configuration of the probed planets. Indeed, the special geometrical configuration of transiting planets offers the detailed study of their atmosphere without the cost of spatially resolving them from their host stars \citep{Winn2010}. The first atmospheric studies of transiting "hot Jupiters" (e.g. \cite{Charbonneau2009, Sing2016}) and "hot Neptunes" (e.g. \cite{Deming2007b, Wakeford2017})  have provided initial glimpses at the atmospheric chemical composition, vertical pressure-temperature profiles, albedos, and circulation patterns of extrasolar worlds (see \cite{Madhu2019} for a recent review). Within the last decade, such atmospheric studies have been extended to smaller and/or cooler planets like the hot rocky worlds 55\,Cnc\,e \citep{Demory2012, Demory2016} and LHS3844\,b \citep{Kreidberg2019} or the more temperate mini-Neptunes GJ1214\,b \citep{Kreidberg2014} and K2-18\,b \citep{Benneke2019}.   

\begin{figure}
    \centering
    \includegraphics[width =1\columnwidth]{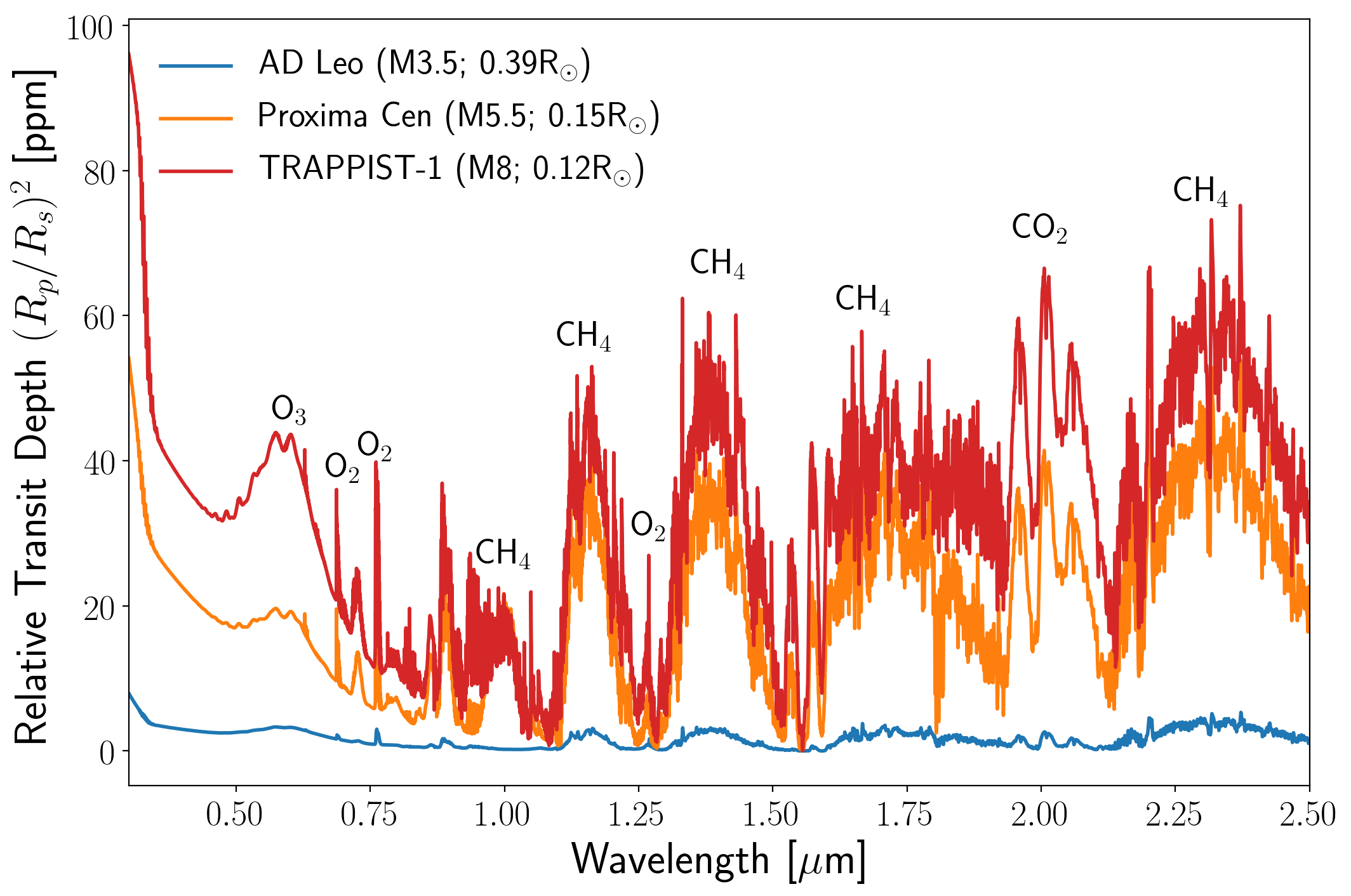}
    \caption{Transmission spectra for three photochemically self-consistent Earth-like planets around
three M dwarfs with different radii. Original references for the transmission spectra simulations are \cite{Segura2005}  and \cite{Schwieterman2016} for AD Leo, \cite{Meadows2018} for Proxima Cen, and Meadows et al. (in prep) for TRAPPIST-1.}
    \label{fig:mspectra}
\end{figure}

Extending further such studies to rocky worlds orbiting within the circumstellar habitable zone (HZ) of their star (e.g. \cite{Kasting1993}) would make it possible to probe the atmospheric and surface conditions of potentially habitable worlds, to explore their atmospheric compositions for chemical disequilibria of biological origins \citep{Schwieterman2018}, and, maybe, to reveal the existence of life beyond our solar system. This extension is not possible with the astronomical facilities currently in operation, including with the Hubble Space Telescope (HST), but it could be initiated from 2021 with the launch of JWST. Thanks to its much improved infrared spectral coverage, sensitivity, and resolution over HST, JWST should in theory be able to probe by eclipse  (i.e. transit or occultation) spectroscopy the atmospheric composition of a transiting temperate planet similar in size and mass to Earth.  Nevertheless,  signal-to-noise considerations result in  the  possibility of such an achievement only for a planet transiting a very nearby (up to 10-20 pc) and very-late-type ($\sim$M5 and later) M-dwarf \citep{Charbonneau2009, Lisa2009, dewit2013}. Indeed, the same planet orbiting a bigger star will produce smaller transmission and emission signals, and this drop is very sharp when one gets away from the bottom of the main-sequence. 
For instance, Figure \ref{fig:mspectra} shows transmission spectra for three photochemically self-consistent Earth-like planets around three different M dwarfs with different radii. The relative transit depth of the prominent near-IR CH$_4$ bands of an Earth-like planet orbiting an M8-type dwarf star is expected to be $\sim$50ppm, but only $\sim$5 ppm for a M3.5V host star. The floor noise of JWST instruments is expected to be of at least 10 ppm \citep{Fauchez2019a}, this `JWST opportunity' is thus clearly limited to the latest-type red dwarfs. This conclusion is strengthened by the habitable zone moving closer to the star as one moves down the main-sequence. For the latest-type red dwarfs, this zone corresponds to orbital periods of only a few days, making possible to stack dozens if not hundreds of eclipse observations within the JWST lifetime, and thus to maximize the final signal-to-noise.  

Such considerations motivated the emergence of several projects aiming to search the nearest M-dwarfs for transits before the launch of JWST, e.g. MEarth \citep{Nutzman2008} or TESS \citep{deming2009}. Among them, the ground-based survey SPECULOOS \citep{Gillon2018, Burdanov2018, Delrez2018b} chose to gamble on `ultracool' (i.e. later than M6) dwarf stars, at the bottom of the main-sequence, despite that most theoretical expectations presented such super-low-mass ($\le 0.1 M_\odot$) stars as unlikely to form short-period rocky planets similar in size to Earth (e.g. \cite{Raymond2007}). While SPECULOOS started officially its operation in 2019 \citep{Jehin2018, Ducrot2019}, it was in fact initiated in 2011 as a prototype survey targeting a limited sample of fifty nearby ultracool dwarfs with the TRAPPIST 60cm robotic telescope in Chile \citep{Gillon2013}. This prototype aimed only to assess the feasibility of SPECULOOS, but it did much better than that. Indeed, in 2015, it detected three transiting Earth-sized planets in orbit around a Jupiter-sized M8-type star 12 parsec away \citep{Gillon2016}. In 2016, the intensive photometric follow-up of the star (renamed TRAPPIST-1) - with a critical contribution of the Spitzer Space Telescope - revealed that it was in fact orbited by seven transiting Earth-sized planets, all of them close or within the HZ of the star \citep{Gillon2017}. With semi-major axes ranging from 1 to 6\% of an astronomical unit, these  planets form a super compact planetary system whose long-term stability is ensured by its resonant architecture \citep{Luger2017a}. While transit photometry brought precise measurements of the planets' sizes  \citep{Delrez2018a}, their strong mutual interactions made possible to infer initial estimates of their masses using the transit timing variations method \citep[TTVs;][]{Agol2005, Holman2005}, the resulting densities suggesting rocky compositions maybe more volatile-rich than Earth's (\cite{Grimm2018}, Agol et al. in prep.). 

The discovery of TRAPPIST-1 was followed by several theoretical studies aiming to assess the potential of JWST to probe the atmospheric composition of its planets \citep{Barstow2016, Morley2017, Batalha2018, Krissansen2018, Wulderlich2019, Fauchez2019a, Lustig-Yaeger2019}. These studies converged on the following picture: assuming  significant atmospheres around the planets, they could be detected by JWST in transmission --and in emission for the two inner planets-- assuming the observation of a number of transits/occultations ranging from less than ten to more than one hundred, depending on the atmospheric properties of the planets (composition, cloud coverage) and the still unknown instrumental performances of JWST. The JWST mission will  thus bring an exciting opportunity to learn about the atmospheric properties of a significant sample of temperate Earth-sized planets, but at the cost of a particularly telescope-time-consuming project that will inevitably face many challenges in all its aspects: telescope time granting, observation scheduling, global strategy, coordination between the different involved teams,  constant reappraisal of the strategy as a function of the actual instrumental limitations and first results, data analysis techniques, complementary space-based and ground-based observations, etc. In the following sections, we review in detail these different challenges, to finally conclude that such an ambitious program would strongly benefit from the support of a community initiative aiming to develop a well-defined sequential structure for the study of the system with JWST and to coordinate on every aspect of its preparation and implementation, both on the observational and theoretical levels. But first we examine the uniqueness of TRAPPIST-1 in the next section. 

\section{On the uniqueness of TRAPPIST-1}

One could argue that it is too early to consider TRAPPIST-1 as a prime target for a large JWST program while better targets in the temperate Earth-sized regime could be about to be found around nearer and/or smaller stars by TESS \citep{Ricker2015} or other transit surveys. For instance, a recent study shows that the collisional absorption signature of O$_2$ could be detected by JWST for a TRAPPIST-1e-like planet, but only if located at less than 5pc \citep{Fauchez2020}. To assess this `better target to come' hypothesis, we cross-matched the Gaia DR2 catalog \citep{Gaia2018} with the 2MASS point-source catalog \citep{2MASS} to build a list of M-dwarfs and L-dwarfs within 40pc (see Annex \ref{annex1}). For each of them, we estimated the basic physical parameters (mass, size, effective temperature), and then computed the value of a metric for JWST transmission spectroscopy in the near-infrared (with NIRSPEC in prism mode) of a putative transiting `Earth-like' planet, i.e. with the same mass, size, and equilibrium temperature ($T_{eq}$) than our planet. We also performed the same exercise for occultation photometry in the mid-infrared (with MIRI in imaging mode) for an Earth-sized planet with an irradiation four times larger that of Earth (as TRAPPIST-1b), at the upper limit of the temperate regime ($T_{eq} \sim 400$K). 

\begin{figure*}
    \centering
    \includegraphics[width =\columnwidth]{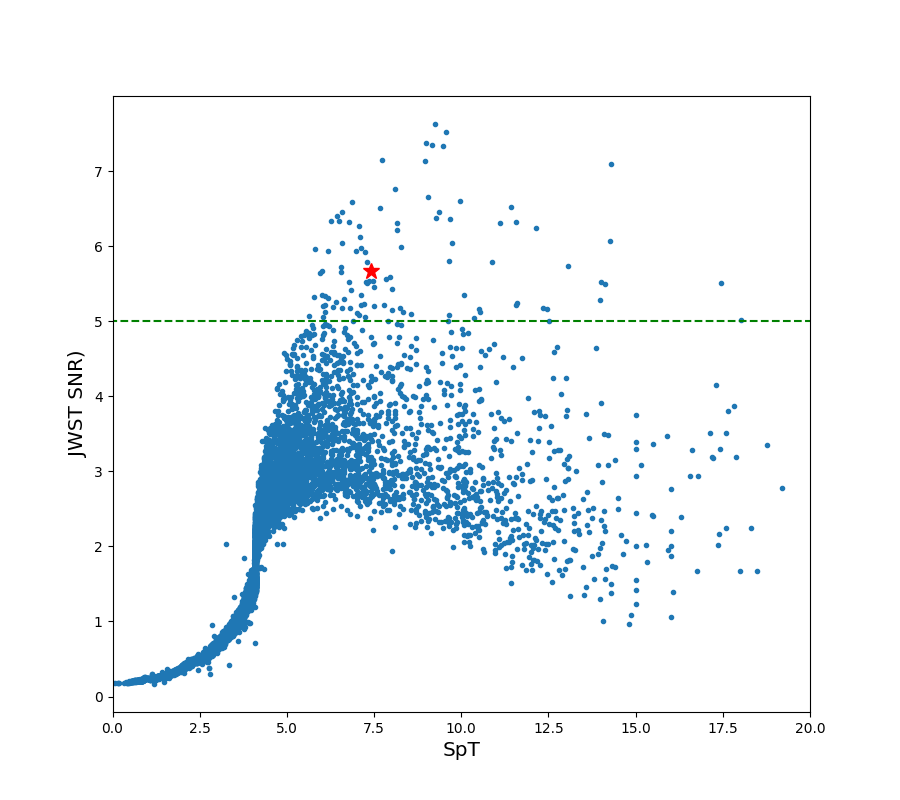}
    \includegraphics[width =\columnwidth]{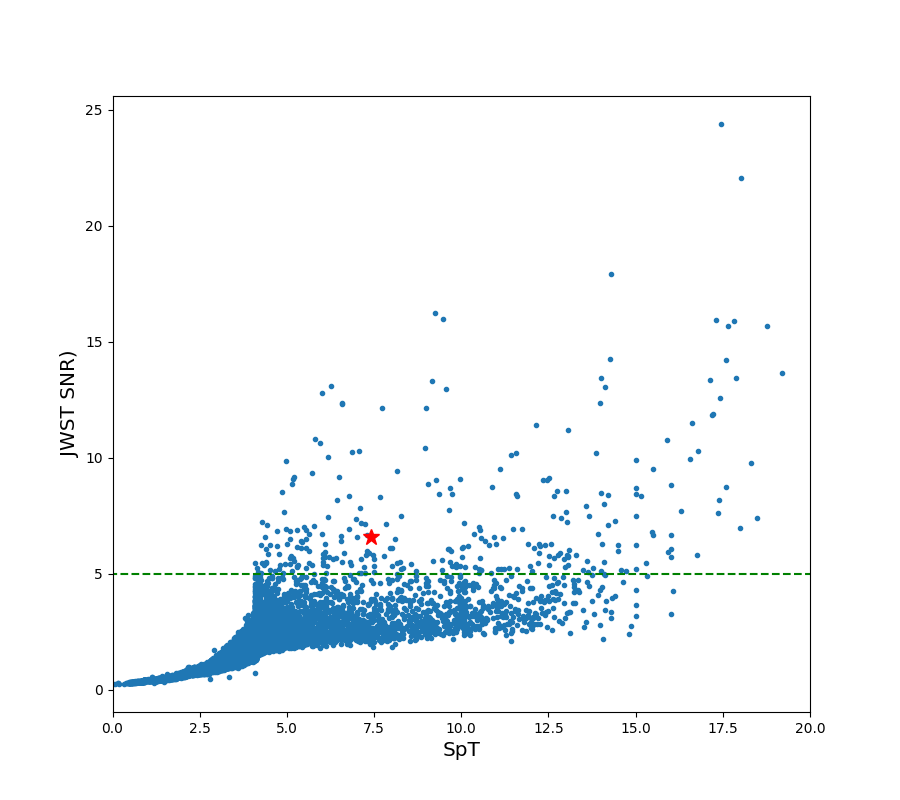}
    \caption{$Left$: Estimated signal-to-noise ratio (SNR) in transmission spectroscopy with JWST/NIRSPEC (assuming an `Earth-like' planet, 200hr of JWST time, and a spectral sampling of 100nm) as a function of spectral type for the 14,163 M- and L-dwarfs within 40pc (see Annex \ref{annex1}). TRAPPIST-1 is shown as a red star. $Right$: same but for occultation photometry with JWST/MIRI (assuming black-body spectra for the star and the planet; a tidally-locked  Earth-sized planet with an irradiation four times larger than Earth, a Bond albedo of 0.1, and an inefficient heat distribution; and 25hr of JWST/MIRI observation in imaging mode within the F1000W ($10 \pm 1 \mu m$) filter).}
    \label{fig:jwst}
\end{figure*}

The details of these analyses are described in Annex \ref{annex1}. Their results are summarized in Fig. \ref{fig:jwst}.  For near-infrared transmission spectroscopy of an `Earth-like' planet, only 44 nearby dwarfs turn out to have a metric larger than TRAPPIST-1. Their spectral types range from M5.5 to L3.5. The mean transit probability of the assumed  planets is $\sim$2.5\%. Except if assuming a particularly high frequency for such planets around ultracool dwarfs, it is thus  likely that TRAPPIST-1d, e, f, and g are the best existing targets for JWST in the `potentially habitable Earth-sized' regime.  Most of these dwarfs ($\sim$70\%) have already been explored for transits by the SPECULOOS transit survey \citep{Gillon2018}, without success except for TRAPPIST-1. 

For mid-infrared occultation photometry of TRAPPIST-1b-like planets, we find that 131 nearby dwarfs have a better metric than TRAPPIST-1, most of them being of later spectral type.  Their mean transit probability is $\sim6\%$. The frequency of short-period Earth-sized planets around ultracool dwarfs is still mostly unconstrained, but except if it is particularly low, there is  a significant probability  that a JWST target at least as good as TRAPPIST-1b could be found soon by SPECULOOS or another ultracool dwarf photometry survey (e.g. \cite{Metchev2019}). Still, if this planet exists, it is unlikely to benefit from all the extraordinary  advantages brought by TRAPPIST-1: \begin{itemize}
    \item Unlike TRAPPIST-1, it could belong to a non-resonant system, as most of the known compact systems \citep{WinnFab2015}. In this case, the precise measurement of the planet's mass should have to rely on the radial velocity (RV) technique instead of the TTV one. Because the host star/brown dwarf will be very faint in the visible, the RV measurements will have to be performed in the near-infrared (NIR). Even with the next-generation high-precision NIR spectrographs like SPIRou \citep{2019MNRAS.488.5114K} or NIRPS \citep{2017SPIE10400E..18W}, reaching the required sub-m.s$^{-1}$ precision level  will be extremely challenging. In fact, such a  level of RV precision has still to be demonstrated in the NIR, and furthermore reaching it could just be  impossible if the NIR RV `jitter' of the host dwarf is significant. 
    \item If by chance the planet belongs to a compact resonant system, it is very likely that, unlike for TRAPPIST-1, its outer planets will not transit. Indeed, the level of coplanarity of the TRAPPIST-1 system is extremely high, and the inclination of its mean  eclipitic is extraordinarily close to 90$^\circ$ ($89.73 \pm 0.09$, \cite{Delrez2018a}).  The probability that another nearby ultracool dwarf system is seen so well edge-on is very small. Even in case of strong TTVs due to a resonant configuration, constraining very accurately the planet's  mass by the TTV technique will be extremely challenging if not all planets of the system transit. 
    \item While the atmospheric characterization of another `warm temperate' ($T_{eq} \sim 400K$) planet would be very interesting, TRAPPIST-1 provides us with the unique opportunity to study not one but seven temperate planets born from the same protoplanetary disk, orbiting the same star, sharing similar masses and sizes, but covering a larger range of irradiation (0.1 to 4 times the Earth's, corresponding to $T_{eq}$ ranging from 170 to 400K). It makes TRAPPIST-1 a unique laboratory to study the impact of the long pre-main sequence phase of ultracool dwarf stars and their strong XUV luminosity,   winds, and flares on the atmospheres and potential habitability of their temperate planets \citep{Airapetian2019}. The comparative study of these seven planets could also enable one to detect a clear outlier planet in the system in terms of atmospheric properties, hinting to something exceptional having affected (e.g. a very energetic impact) or still affecting (e.g. super-volcanism, life) the planet. 
    \item The densities measured for TRAPPIST-1 planets suggest rocky compositions enriched in  volatile compounds  relative  to Earth \citep{Grimm2018}, which favors the existence of significant atmospheres around them, as large volatile reservoirs could efficiently replenish these atmospheres in case of strong erosion by XUV photons and high-velocity ions from the star.
    \item TRAPPIST-1 is an old  star \citep{Burgasser2017} which, when compared to most ultracool dwarfs, rotates probably  more slowly \citep{Luger2017a} and is less active and quieter in optical and NIR photometry \citep{Delrez2018a, Ducrot2018}. Notably, the high stability of the quiescent flux of  the star in the infrared \citep{Delrez2018a} combined to the low frequency of its flares (one flare every 1-2d, one superflare every $\sim$6 months) is a big advantage for NIR/MIR eclipse spectrophotometry with JWST. 
    \item Several studies \citep{Morris2018, Delrez2018a, Ducrot2018, Wakeford2019} failed to detect for TRAPPIST-1 traces of a photosphere inhomogeneous and variable enough  to significantly affect JWST transmission spectra of its planets through spot-crossing events \citep{Espinoza2019} and/or through the so-called `transit light source' effect originating from a difference in mean temperature between the mean photosphere of the star and the stellar chord transited by the planet \citep{Rackham2018}. 
\end{itemize} 

All in one, we can thus safely conclude that the `better target to come’ hypothesis is extremely unlikely.

\section{The star TRAPPIST-1}\label{sec:star}

To obtain the most precise and accurate measurements of the TRAPPIST-1 planets, we need to properly characterize its low-mass host star. Numerous studies to date have been able to firmly constrain many of the star's physical properties, including average density (from the planet transits), bolometric luminosity, mass, radius, and effective temperature; and have ruled out the presence of wide stellar or brown dwarf companions \citep{2019ApJ...886..131G,2018ApJ...853...30V,Gillon2017,F2015,Luger2017a,2016ApJ...829L...2H,2018ApJ...858...55P}. Analysis of the high-energy output of TRAPPIST-1, including its degree of magnetic activity and flare frequency, have been critical for interpreting its atmosphere, structure, and evolution. Like other late-M dwarfs, TRAPPIST-1  produces numerous strong flares \citep{2018ApJ...858...55P}, has persistent H$\alpha$ ($L_{H\alpha}/L_{\rm bol} = 2.5-4.0\times10^{-5}$) and X-ray emission ($L_{X}/L_{\rm bol} = 2-4\times10^{-4}$), and may possess a strong surface magnetic field (600$^{+200}_{-400}$~G; \citealt{Reiners2010,2015ApJS..220...18B,Wheatley2017}). While in some of these measures TRAPPIST-1 is less active than other M8 dwarfs \citep{Burgasser2017}, its magnetic activity is sufficient to influence both its physical structure \citep{2018ApJ...869..149M} and the potential habitability of its planets (\cite{Vida2017, Wheatley2017, Garraffo2017}, see \S \ref{sec:atmo}). 

Despite these numerous studies, the age, rotation period, magnetospheric structure, and detailed abundances are all relatively poorly constrained \citep{Burgasser2017, 2019ApJ...886..131G,2007ApJ...656.1121R}, and there remain mysteries as to its size and spectral properties. TRAPPIST-1's age remains highly uncertain as its spectral properties and kinematics give conflicting indicators \citep{Burgasser2017, 2019ApJ...886..131G}, while standard age-dating metrics of rotation and activity are largely unreliable for ultracool dwarfs. Remarkably, despite many monitoring studies, even the star's rotation period remains uncertain \citep{2017NatAs...1E.129L,Delrez2018a}, and there may be conflicting assessments as to the degree and thermal properties of magnetic spotting \citep{2018ApJ...863L..32M, Wakeford2019}.  The star also displays a novel and unique pattern of flares which appear to be correlated with quasi-periodic bright spots \citep{2018ApJ...863L..32M}.  These measures play a significant role in determining why TRAPPIST-1 has a lower average density than predicted by evolutionary models, with the roles of magnetism and metallicity both in play, despite uncertain measurements of both \citep{Burgasser2017,2018ApJ...853...30V,2018ApJ...869..149M}. Continued study of TRAPPIST-1 from the ground and from space will be essential to addressing these mysteries, to better understanding the formation, evolution, and environment of its planetary system, and to interpreting and putting into perspective the JWST results.  \\

\section{Dynamics, masses, and compositions of the TRAPPIST-1 planets} \label{sec:dyn}

The TRAPPIST-1 planets lie in (or close to) a series of 3-body Laplace resonances \citep{Luger2017a}, with an extremely co-planar configuration \citep{Luger2017b}. This dynamical configuration is characteristic of a small number of multi-planet systems found with Kepler \citep[e.g. Kepler-80,][]{MacDonald2016}, and indicates that the planets' orbits migrated inwards, capturing into resonance \citep{Papaloizou2017}.  The resonant configuration also means that the planets are in close proximity to one another, which causes strong planet-planet perturbations, leading to significant TTVs.

The TTVs in this system turn out to yield strong constraints on the masses and orbits of the planets  \citep{Gillon2017, Grimm2018}. The small star causes the planets to be large relative to the star, giving dual advantages:  the large mass-ratios of the planets to the star causes stronger TTVs (relative to the planet orbital periods), and the large depths of transit yield precise timing measurements.   Transit-timing of TRAPPIST-1 planets yielded the first constraints on the masses \emph{and} radii of Earth-sized planets near the same irradiation as Earth \citep{Grimm2018}.  A series of large Spitzer and ground-based programs to measure the times of transit for all seven planets has just concluded, and the data are being analyzed.  The results of this analysis will yield more precise constraints upon the masses of the planets, as well as an improved forecast of the transit times of the planets for the future.  Both masses and ephemerides are necessary for more accurate planning of JWST observations:  the densities of the planets constrain possible bulk compositions; the masses are of prime importance for the robust characterization of the planets' atmospheres \citep{Batalha2019}; the surface gravities of the planets may help to forecast the atmospheric scale-heights, and hence the spectroscopic transit depths; and the transit times will help to plan JWST observations. 

Initial TTV analyses yielded low densities for the planets relative to an Earth-like composition (with the exception of planet e).  The low densities indicate several possible scenarios for the planets' compositions:  a lower fraction of mass in an iron core;  a higher abundance of water \citep{Unterborn2018a,Unterborn2018b,Dorn2018}; or perhaps a thick atmosphere.  Refined measurement of the planet densities (Agol et al. in prep.) will help to distinguish between these possibilities.

\begin{figure*}
    \centering
    \includegraphics[width=2\columnwidth]{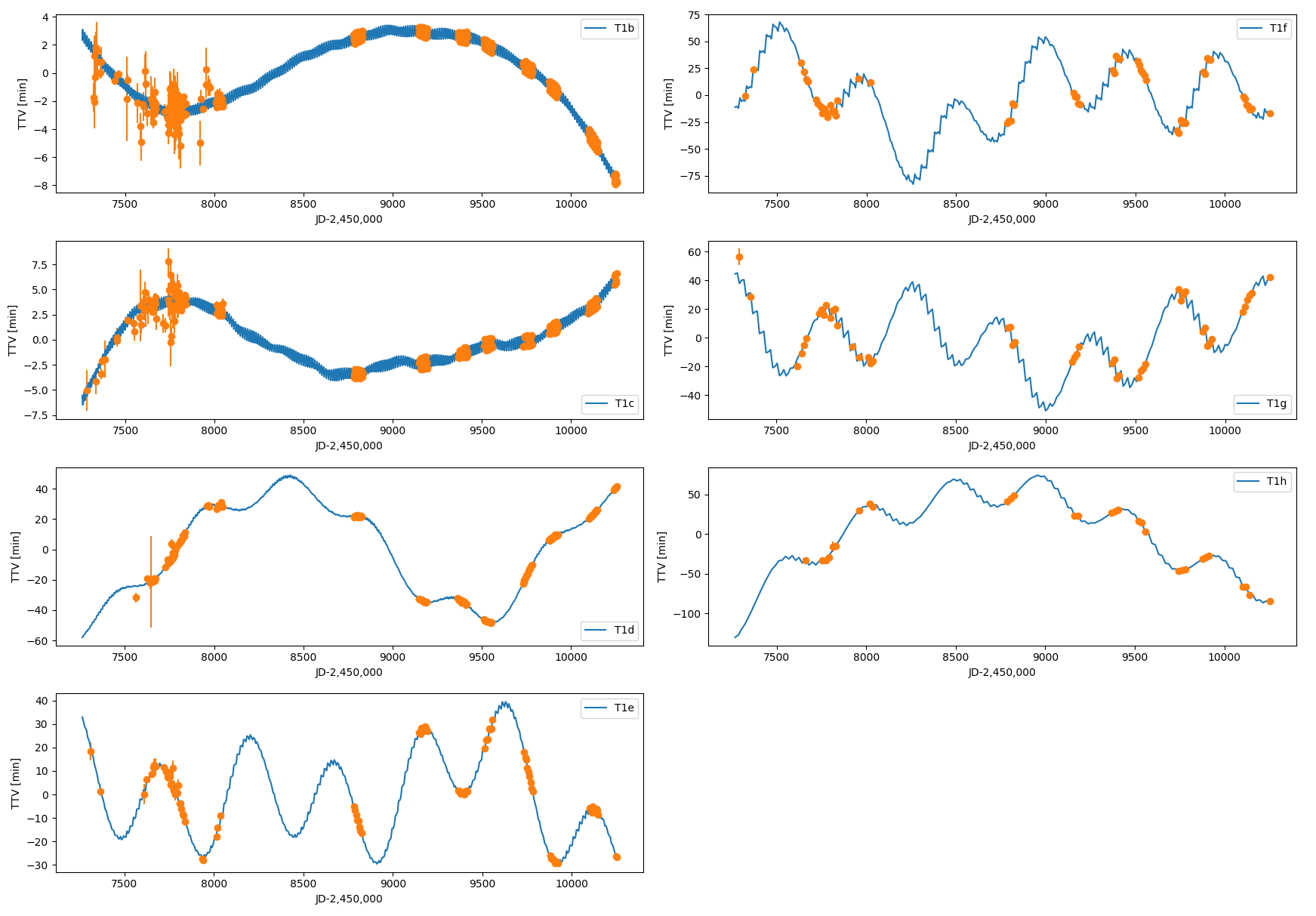}
    \caption{Simulated TTVs for JWST assuming \emph{all} transits were observed for a 5-year mission.}
    \label{fig:JWST_TTV}
\end{figure*}

With JWST, numerous observations of the transit of each planet will be necessary to measure their transmission spectra. With each transit comes a transit time, impact parameter and duration measurement;  together these will yield further constraints upon the dynamics of the planets, including their masses. The timing precisions of the transits will be of order of $\approx$ 1 second, and may be limited by stellar variability (B. Morris, priv.\ comm.).  If \emph{every} accessible transit  were measured with JWST over 5 years (about 1000 transits; Figure \ref{fig:JWST_TTV}), then the masses of the planets may be constrained relative to the star with a precision of parts-per-thousand, or better (E. Agol, priv.\ comm.).  Of course a smaller number of transits will be observed, but nevertheless this indicates the  potential power of JWST for constraining the dynamics of these planets, including the masses, which will yield better constraints upon the compositions.  JWST will also yield  more precise measurements of the transit durations, which will further constrain the density of the star and the coplanarity of the system \citep{Seager2003}.  An improved stellar density is needed for better constraining the densities of the planets as TTVs yield the \emph{mass ratio} of the planets to the star, while the transit depths yield the \emph{radius ratio} of the planets to the star \citep{Agol2017}.

Some remaining questions and puzzles exist in the dynamical analysis of the TRAPPIST-1 system.  The free eccentricities measured from the TTVs appear to be large for the inner planets, $\approx 1$\%, and the longitudes of periastron of the inner two planets are aligned rather than anti-aligned, at odds with the expectation based upon tidal or viscous damping \citep{Papaloizou2017}; although this discrepancy has low significance.  This result may be due to the weak sensitivity of the TTVs to eccentricity, and the constraints on eccentricity should improve with JWST.  The existence of planets beyond the seven detected transiting planets might affect their TTVs;  JWST will give the opportunity to search for additional planets.  The rotational state of the planets may affect the dynamics:  if the planets are tidally synchronized, then perhaps the dynamical state may be influenced by the interior structure of the inner planet(s)  \citep{Mardling2007,Batygin2009}.  Another intriguing possibility  is the existence of orbital obliquities driven by resonant interactions, akin to a  Cassini state \citep{Millholland2019}.  Such dynamical configurations might be probed with  transit timing \citep{Ragozzine2009}.  A measurement of the $k_2$ of the inner planet would help to constrain its composition (\cite{Baumeister2019}; Bolmont et al. submitted).  More theoretical work is needed to investigate these possibilities, and their potential detectability with JWST.

\section{On the possibility of atmospheres around the TRAPPIST-1 planets}
\subsection{Theoretical expectations}
\label{sec:atmo}

The short orbital distances of TRAPPIST-1 planets combined to the large high-energy luminosity (XUV photons + winds) of their host star suggest that atmospheric escape should have played an important role in the evolution of their atmospheres \citep{Vida2017, Garraffo2017, Bourrier2017, Roettenbacher2017, Cohen2018, Tilley2019, Peacock2019, Fraschetti2019, Rodriguez2019}. Should the planets have reached their final masses before gas disc dispersal, core-accretion models imply that planets b, c, f \& g were massive enough to accrete primordial hydrogen-dominated atmospheres that represented a few percent of the planet's mass, unlike d, e \& h \citep[e.g.][]{Lee2015}. Primordial evolution models for planets around M dwarfs \citep[e.g.][]{Owen2016} suggest that while b, c, d, e \& h would have been able to lose such small amounts of H/He, it is not the case for planets f \& g. Given that the measured densities \citep{Grimm2018} and atmospheric reconnaissance observations rule out large hydrogen-dominated atmospheres \citep{deWit2018} including for these two later planets, it suggests that the TRAPPIST-1 planets did not accrete significant hydrogen-dominated atmospheres from their parent protoplanetary disc.   

Given the measured densities of TRAPPIST-1 planets favor some fraction of volatiles and the location of  some of them within the HZ, some authors have studied the loss of steam-dominated atmospheres. This is a particular pertinent questions as even though some of the planets may reside in the HZ now, the long pre-main-sequence lifetimes of late-type M dwarfs imply that the planets may have spend several 100~Myr in a runaway greenhouse state \citep{Luger2015}. In  such a state, greenhouse water can reach the thermosphere, where it can be photo-dissociated and undergo escape \citep[e.g.][]{Kasting1988}. Hydrogen atoms are easier to lose; however, heavier elements such as oxygen can be ``dragged'' along too. However, if the oxygen remains a large abiotic oxygen-dominated atmosphere can be produced \citep[e.g.][]{Luger2015}. Both \citet{Luger2015} and \citet{Bolmont2017} have attempted to model the the loss of water from TRAPPIST-1-like planets. While these studies show that the planets are vulnerable to water-loss, they disagree on the extent of volatiles lost on the final outcomes of TRAPPIST-1-like planets, indicating that the physics of water escape is not a solved problem and requires further study and observation. Finally, understanding whether or not the TRAPPIST-1 planets can retain a heavy element dominated atmosphere requires more theoretical work. There are no quantitative predictions for what atmospheric escape from heavy element dominated (e.g. CO$_2$, H$_2$O) atmospheres would look like over the lifetimes of the TRAPPIST-1 planets.  This includes thermal escape models and non-thermal escape models \citep[e.g.][]{Ribas2016, Dong2018}. Ultimately, observations of the presence of an atmosphere on the TRAPPIST-1 planets and their compositions will be of fundamental importance to constraining atmospheric evolution and escape models.  

\subsection{First observational constraints from HST}

When the planets were detected, it was directly realized that HST and its Wide Field Camera 3 instrument should be sensitive enough to reveal primordial cloud-free hydrogen-dominated atmospheres. This observation motivated the HST programs 14500, 14873 \& 15304 (PI: J. de Wit) that first ruled out this atmospheric scenario for planets b and c in May 2016 \citep{deWit2016}, and then for planets d, e, f, and g over December 2016 and January 2017 \citep{deWit2018, Wakeford2019}. The last planet of the system, planet h, has now been observed twice with HST which led to the same negative result (Wakeford et al., in prep).  Current observations are thus ruling out the presence of cloud-free hydrogen-dominated atmospheres for all planets of the TRAPPIST-1 system. However, high-altitude clouds and hazes are not expected in hydrogen-dominated atmospheres under such a range of isolation \citep{Hu2014, Morley2015}, which, when combined to the theoretical arguments mentioned in Sec. 5.1, make primordial atmospheres around TRAPPIST-1 planets unlikely.

Subsequently, HST’s unique UV capabilities were leveraged to search for the presence of hydrogen exospheres around the planets (program 14900 \& 15304). The analysis of the first part of these data revealed to be complicated by the extreme faintness and the variability of the star in Lyman-$\alpha$ \citep{Bourrier2017, Bourrier2017b}. Finally, the observations obtained for all seven planets failed to detect hydrogen exospheres around all planets, thereby placing an upper limit on their current atmospheric escape rates (manuscript in prep.). 

In summary, the atmospheric reconnaissance of the system with HST has left three hypotheses for each planet: (1) no significant atmosphere, (2) a cloudy hydrogen-dominated atmosphere (unlikely), and (3) a compact atmosphere. JWST should be able to discriminate between these three scenarios.  

\section{TRAPPIST-1 under the eye of JWST}

\subsection{Transmission spectroscopy}

The first step in characterizing significantly compact atmospheres around the TRAPPIST-1 planets is of course to detect them unambiguously. Low-resolution transit spectroscopy with JWST is likely to be the best method for atmospheric detection for all planets except the two inner ones. Planets b and c could have dayside thermal emissions large enough to be detected through the observation of a few of their occultations with MIRI, which could bring key complementary insights into the presence of an atmosphere around them \citep{Koll2019}.  Transit spectroscopy leverages the long slant path through the exoplanetary atmosphere to produce detectable molecular absorption signatures even if the atmosphere is thin, or largely masked by clouds.  Moreover, the host star is bright in the infrared (K = 10.3), providing high signal-to-noise ratios in, e.g., the wide bandpass (0.6-5$\mu$m) of the NIRSpec Prism mode. 

\begin{figure}
    \includegraphics[width=0.45\textwidth]{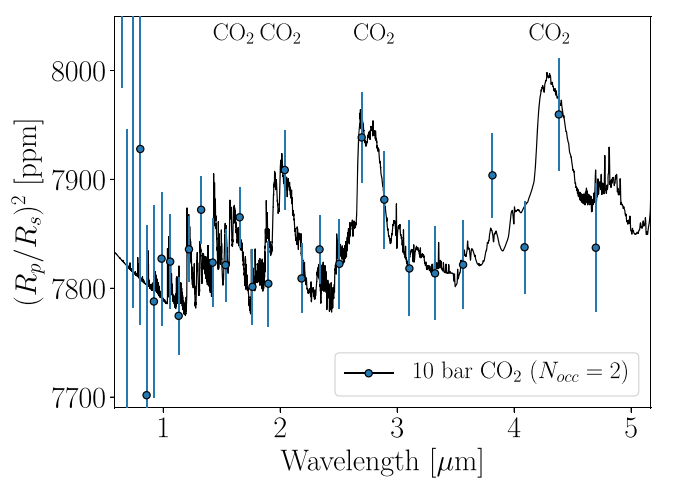}
    \caption{Example of a NIRSpec prism transit spectrum of TRAPPIST-1b, from \citet{Lustig-Yaeger2019}.  This is for two transits, binned to a spectral resolving power of 8.  Note the detections of the strong carbon dioxide bands.}
    \label{fig:tran_spec1}
\end{figure}

\begin{figure}
    \includegraphics[width=0.45\textwidth]{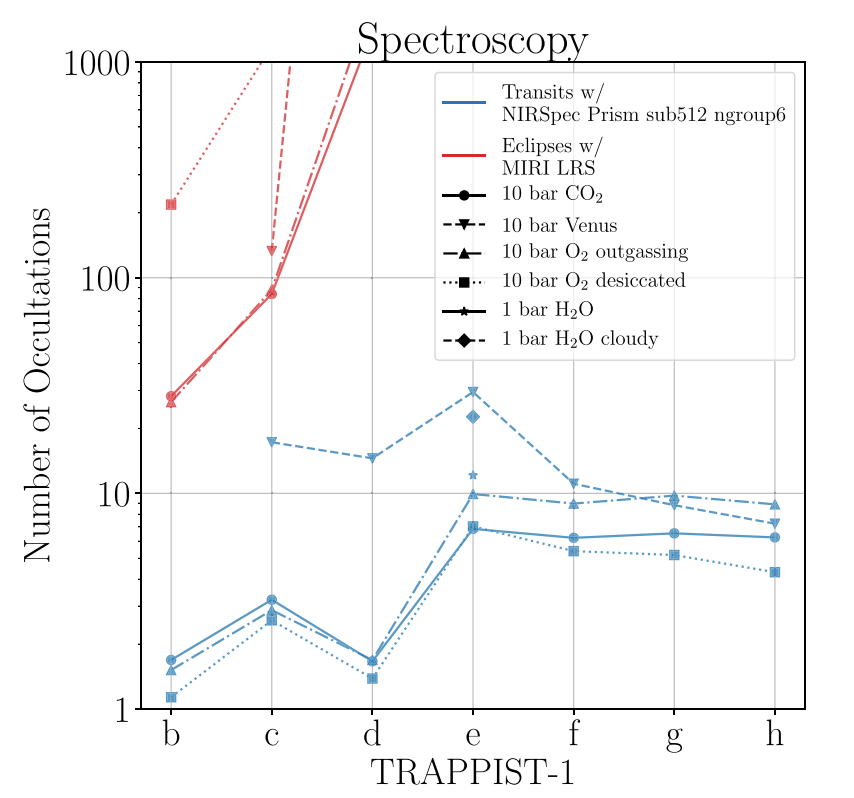}
    \caption{Total number of eclipses needed to detect atmospheres of various compositions using occultation emission spectroscopy with MIRI LRS (red lines) and 
    transit transmission spectroscopy with NIRSpec prism (blue lines). From \citet{Lustig-Yaeger2019}. }
    \label{fig:tran_spec2}
\end{figure}

An in-depth study of how the atmospheres of the TRAPPIST-1 planets could be detected and characterized was reported by \citet{Lustig-Yaeger2019}.   It confirmed that transit spectroscopy is optimum for atmospheric detection, particularly for temperate and cool planets.  Figure~\ref{fig:tran_spec1} shows a simulated detection of the atmosphere of planet b in two transits, assuming it has a clear atmosphere with 10 bars of carbon dioxide.  Figure~\ref{fig:tran_spec2} shows the number of transits or occultations needed for atmospheric detection of each planet in the system using both occultation spectroscopy with MIRI LRS and transit spectroscopy with the NIRSpec prism.  The atmospheres of planets b and c are most easily detected, and most of the modeled atmospheres and planets can be detected in less than 10 transits (including clear, cloudy, and atmospheres containing Venus-like sulphuric acid aerosols). 

Three TRAPPIST-1 planets will be targeted with transmission spectroscopy by JWST GTO programs, namely planets d (two transits with NIRSPEC, program 1201, PI: D. Lafreni\`ere), e (four transits with NIRSPEC, program 1331, PI: N. Lewis) and f (five transits with NIRISS, program 1201). It is desirable to assess the potential of these GTO observations and, as of Cycle 1, to complement them with more transit observations, if required, and to perform similar observations for the four other planets. Indeed, the detailed atmospheric characterization of some of these planets could require the observation of dozens of eclipses (e.g. \cite{Lustig-Yaeger2019}), while JWST's mission lifetime could be as short as 5.5 yr. Furthermore, TRAPPIST-1 lies near the ecliptic and  is only observable with JWST during two visibility windows of only $\sim$50 days per year, limiting the  number of transits or occultations observable per year to at most 65, for b, and down to 5 for h. It is thus crucial to start the atmospheric characterization of the seven planets as of Cycle 1, without waiting for the results for one given planet and/or one instrumental set-up to initiate more observations. 


\subsection{Emission + phase curve photometry}

Observations of thermal emission from terrestrial planets -- either through occultations  or full-orbit phase curves -- are a powerful probe of atmospheric properties, complementary to transmission spectroscopy. Thermal emission measurements constrain the planet's global temperature structure, which is highly sensitive to the surface pressure and chemical composition of the atmosphere.

For tidally locked planets (as expected for the TRAPPIST-1 system), higher surface pressures lead to more efficient heat circulation from day-side to night-side, evening out the temperature on the two hemispheres. Conversely, the day-side temperature is maximized if no atmosphere is present. The occultation depth or phase curve amplitude can therefore be used as a proxy for the planet's surface pressure \citep[e.g.][]{Seager2009,Selsis2011, Koll2016, Mansfield2019,Koll2019}. 
This approach was recently tested with Spitzer observations of the 1.3 $R_\oplus$ planet LHS~3844b, which revealed such a large day-side temperature that thick atmospheres are ruled out, and the planet is most likely a bare rock \citep{Kreidberg2019}

This result motivates thermal emission measurements for a larger sample of terrestrial planets, over a wider range of equilibrium temperatures \citep[LHS~3844b is the hottest known terrestrial planet orbiting an M-dwarf;][]{Vanderspek2019}. Atmospheres on cooler planets are expected to be less vulnerable to atmospheric escape, and are more likely to maintain stable secondary atmospheres \citep[e.g][]{Bolmont2017}.    The TRAPPIST-1 system has the coolest known terrestrial planets accessible for thermal emission observations (all below 400 Kelvin), so it provides an ideal environment to explore the existence of atmospheres over the widest possible range of isolation.

The innermost two TRAPPIST-1 planets (b and c) are strong candidates for occultation observations. Thermal emission from plausible model atmospheres can be detected in some MIRI photometric filters at $5\sigma$ confidence with 5 - 10 occultation observations per filter \citep{Morley2017}. Additional occultations can reveal the atmospheric composition in detail; conversely, if no atmosphere is present, occultation spectroscopy can potentially distinguish between different rocky surface compositions \citep{Hu2012, Lustig-Yaeger2019}.

Thermal emission measurements have great potential to yield interesting scientific results regardless of the nature of the planet (in particular, whether it has an atmosphere). If an atmosphere is present, occultations and phase curves can probe the atmospheric circulation and chemical composition. Thermal emission is less sensitive to high-altitude clouds or haze that can mute the amplitude of spectral features in transmission spectra \citep{Fortney2005}. TRAPPIST-1b is already targeted by  JWST/MIRI GTO program, with ten occultations planned, five within (program 1177, PI: T. Greene) and five outside (program 1279, PI: P.-O. Lagage) the CO$_2$ absorption band at 15 $\mu$m. It is desirable to perform similar observations for planet c. 

\section{Complementary observations}

While TRAPPIST-1 has been observed intensively from the ground and from space since the discovery of its planets, it is necessary to pursue and even strengthen this effort to maximize the scientific return of its upcoming JWST observations. 

The accurate and precise measurements of the masses of the planets is an absolute prerequisite to the derivation of strong inferences on their atmospheric and surface properties from JWST spectrophotometric data \citep{Batalha2019}. The TTV method has already enabled the measurement of the planet masses with great precision, but the analysis of the current transit timing dataset is still somewhat hindered by degeneracies and uncertainties, and it is highly  desirable to pursue an intense photometric monitoring of the transits of the seven planets to improve further our understanding of the dynamics of the system, e.g. to detect a possible outer eighth planet, to quantify the effect of tides on the dynamics of the inner planets, to reveal possible misalignments between the planets’ orbits, etc, and to improve the determination of the planet’s masses in the process (see  \S \ref{sec:dyn}). Furthermore, pursuing the photometric monitoring of the  transits is required to produce the most accurate possible ephemeris for the JWST transit observations. So far, Spitzer has played a key role in this respect with a total of nearly 200 transits observed at high photometric precision \citep{Gillon2017, Delrez2018a}. Ground-based telescopes have also brought a significant contribution (e.g. \cite{Ducrot2018, Burdanov2019}), but now that the Spitzer extended mission is ending, this contribution has to strengthen. Several ground-based facilities are already heavily involved in this transit monitoring: the eight robotic telescopes participating to the SPECULOOS survey (five in Chile, one in Tenerife, one in Mexico, and one in Morocco), UKIRT (Hawaii), and the Liverpool Telescope (La Palma). It is desirable to extend this monitoring to other facilities at other longitudes, to observe as many transits as possible during the visibility period of the star (from May to December). 

Now that several high-resolution ($\lambda/\Delta\lambda$ $\approx$ 10$^5$) near-IR spectrographs optimized to reach high Doppler precision are --or are close to be-- operational, e.g. SPIRou \citep{2019MNRAS.488.5114K} and NIRPS \citep{2017SPIE10400E..18W}, these could be used to explore the system for more outer planets, but also to robustly measure the stellar $v\sin{i}$ and test for spin-orbit alignment (both potentially achievable using the Rossiter-McLaughin effect).  Synoptic monitoring of TRAPPIST-1 from X-ray to radio wavelengths will improve our understanding of the physical and magnetospheric structure of the star, as well as our ability to characterize the magnetic environment in which the planets reside, and complement simulations of the stellar magnetosphere and radiation environment \citep[e.g.][]{Fraschetti2019,Cohen2018,Garraffo2017}. High-precision, high-resolution spectroscopy  is needed to further reduce current uncertainties on the radius and mass of the star \citep{Mann2019}. These data will also allow us to robustly measure the star's photospheric abundances, which sets a baseline for atmospheric and evolutionary modeling, as well as the composition of the system's natal planet-forming disk. The existing and acquired data will need to be matched to more advanced atmosphere, structure, and evolution models, which take into account non-equilibrium chemistry, magnetic field generation and energetics, magnetospheric structure and flare generation, star-disk-planet interactions, and angular momentum evolution across the system. Synchronizing these observations with JWST measurements will also enable us to put the later into appropriate context, as the inferred properties of the planets are intimately coupled to the current state of this persistently active star.

The combination of multi-epoch high-resolution spectroscopy, obtained for radial velocity measurements or other purposes, and long-term photometric monitoring offers a powerful approach to constraining the photospheric heterogeneity of TRAPPIST-1 and its potential impact on transmission spectra from the system. Photometric brightness variations track relative changes in the projected covering fraction of active regions but lack the necessary reference for establishing the absolute covering fractions if the star is persistently active \citep{Rackham2018}. However, stellar spectral decomposition, which identifies the unique constituents of a full-disk stellar spectrum (e.g., immaculate photosphere, spots, and faculae) by focusing on spectral features that must originate in different temperature regions or that vary in time, provides both the unique spectral components and their projection-weighted covering fractions \citep{Neff1995}. In the context provided by the high-resolution spectra, photometric monitoring can then be used to track absolute changes in active region coverage \citep{Gully-Santiago2017}. In the case of TRAPPIST-1, studies have suggested both two- and three-component models for the photosphere of TRAPPIST-1 (\cite{Morris2018, Zhang2018, Wakeford2019}. A coordinated effort between community members pursing high-resolution spectra and photometric monitoring of TRAPPIST-1 will be useful for uniquely determining the photospheric components present on the star, tracing their evolution (if any), and identifying the photospheric heterogeneity of the star at the time of JWST transit observations and potential for stellar spectral contamination in transmission spectra.

The high sensitivity and angular resolution of JWST will be essential to push the limits for resolved companions, including potential outer giant planets that may have played in role in shaping and/or stabilizing the compact hierarchical configuration of the system.

\section{Modeling the JWST TRAPPIST-1 measurements}

\begin{figure}
    \includegraphics[width=0.5\textwidth]{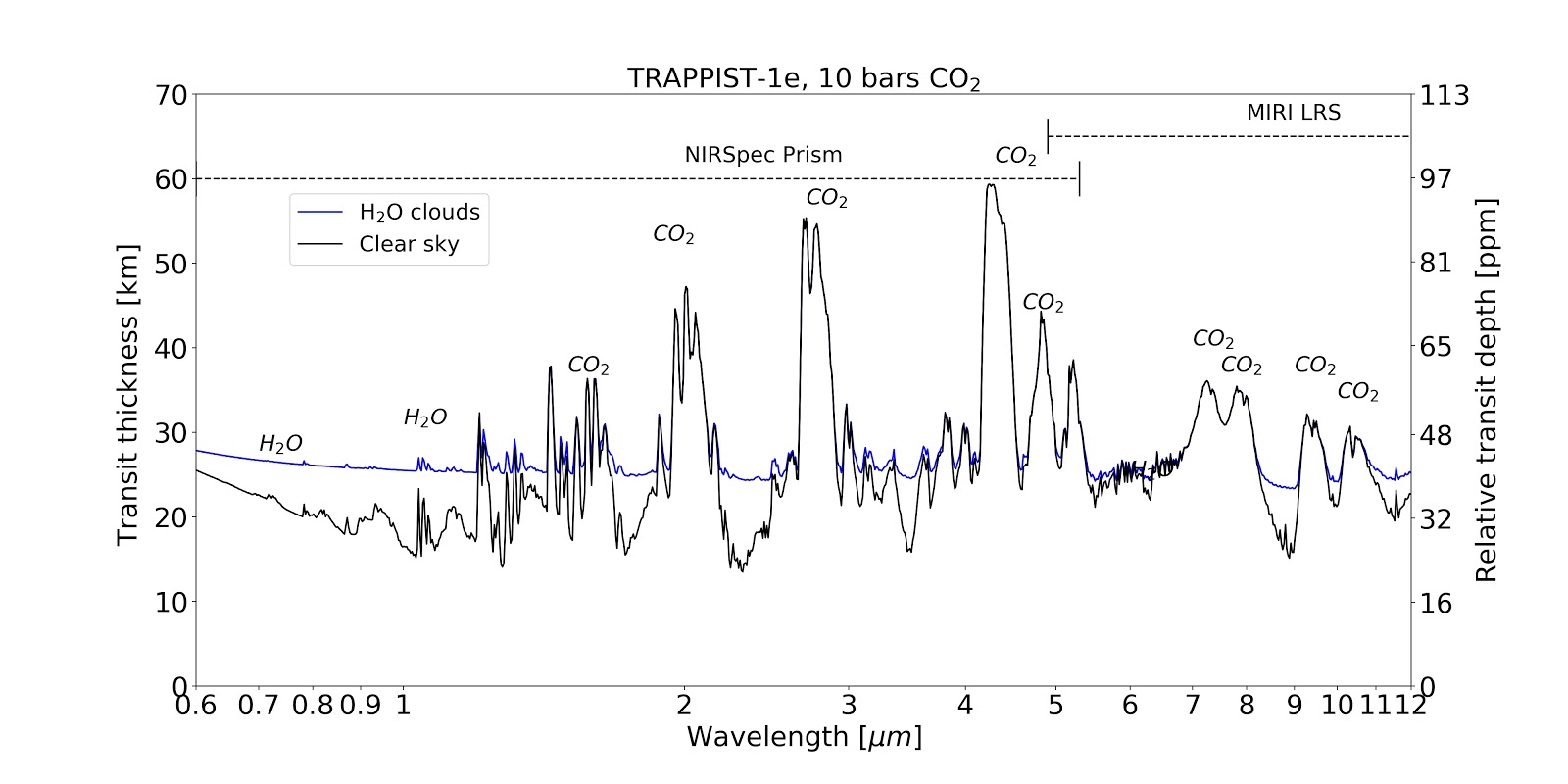}
    \caption{JWST NIRSpec Prism and MIRI transmission spectra (R=300) simulated by the Planetary Spectrum Generator \citep{Villanueva_2018} for TRAPPIST-1e, fully covered by oceans, with 10 bars of surface pressure of CO$_2$ \citep{Fauchez2019a}.  The figure illustrates the effect of H$_2$O clouds, which raise the continuum level up to $\sim$35 kilometers above the surface, flattening the H$_2$O lines and reducing the relative transit depth (or atmospheric thickness) of other species.}
    \label{fig:spectral_models}
\end{figure}

During the last decade, the exoplanet community has achieved significant progress in producing atmospheric models able to capture the expected diversity of composition, dynamical regimes and observable signatures (e.g. Fig.~\ref{fig:spectral_models}). Gathering a comprehensive suite of existing and complementary modeling tools, would insure the best possible preparation, optimisation and eventually exploitation of TRAPPIST-1 observations with JWST.  1D models - that assume an average vertical structure and composition for the whole planet - would allow us an extensive exploration of the parameter space in order to identify the most promising features to search for.  
These models can include detailed physics and are readily coupled to photochemical models that take into account the impact of the star's spectrum on planetary composition.  They also have the versatility to explore a wide range of terrestrial atmospheric compositions that are very different to those of a modern Earth-like planet, and that could potentially reveal signs of ocean loss, interior outgazing or current habitability in the TRAPPIST-1 planetary environments \citep{lincowski2018evolved}.
3D hydrodynamics codes (GCMs: Global Climate Models) can then refine the atmospheric structure and cloud distribution by accounting for the synchronous (or quasi-synchronous) rotation and the resulting day-night thermal contrast, which can dramatically affect observables \citep{Wordsworth_2011, Yang_2013, Shields_2013, Turbet_2016, Kopparapu_2017, Fujii_2017, Haqq_Misra_2018, Way_2018, Fauchez2019a, Wolf_2019, Komaceck_2019, Suissa_2020}.
The amplitude of infrared phase-curves is controlled by the day-night temperature difference at the altitude of emission \citep{Selsis2011, Wolf_2019}.
Transit spectroscopy is also affected by the 3D structure of the atmosphere as the region around the terminator sounded by the method is usually large enough to mix information from both the day and night hemispheres \citep{Caldas_2019}. 
Different GCMs have their own strengths and weaknesses relative to, e.g., the treatment of radiative transfer, convection, or photochemistry, the modeling of the atmosphere-ocean exchanges, the adopted spatial resolution, or the computational speed. These differences can lead to different climate predictions and thus different estimation of the detectability and characterization of exoplanet atmospheres.  Intercomparisons between the models within the community stimulate rapid progress and strengthen observation choices and interpretation of future data \citep{Yang_2019, Fauchez2019a}. The TRAPPIST Habitable Atmosphere Intercomparison (THAI, \cite{Fauchez2019b}) uses TRAPPIST-1e as a benchmark to intercompare four GCMs (LMDG, ROCKE-3D, ExoCAM, UM). THAI includes four test cases: two land planets composed of a modern Earth-like composition (N2 and 400 ppm of CO2 ) and a pure CO2 atmosphere, respectively, and two aqua planets with the same two atmospheric compositions. The differences between the model outputs are compared and their impact on the simulated spectra is evaluated in the context of JWST (and future telescopes) observations (Fauchez et al. in prep). Groups working with GCMs have developed tools to generate synthetic observables accounting for the 3D atmospheric structure for all observing modes considered for TRAPPIST-1 planets (transmission - emission -  at any orbital phase) and for all resolutions/bandwidths of JWST instruments. By providing synthetic observations of known virtual atmospheres, combined with JWST observation simulators, we could test and select retrieval procedures.

\section{On the possibility to detect biosignatures on TRAPPIST-1 with JWST}

JWST will allow us to explore terrestrial exoplanet atmospheres to understand their composition, and in doing so, provide the first opportunity to search for biosignature gases in a handful of terrestrial exoplanet atmospheres. The TRAPPIST-1 system may provide the best targets for this search, which will likely be challenging. Detecting biosignatures will only be feasible if care is taken to coordinate JWST observations that maximize the chances of success over the mission duration.  Numerous studies have examined the potential detectability of oxygen (O$_2$) and its photochemical byproduct ozone (O$_3$) for JWST observations of the TRAPPIST-1 planets (e.g. \cite{Barstow2016, Wulderlich2019, Lustig-Yaeger2019, Fauchez2019a, Hu2020}). A consensus is emerging that these classic signs of oxygenic photosynthesis may require a substantial number of transits be co-added for detection (\cite{Wulderlich2019, Lustig-Yaeger2019, Fauchez2019a}). Indeed, O$_3$ may only be weakly detected in 100 transits with MIRI LRS (\cite{Wulderlich2019, Lustig-Yaeger2019}), while O$_2$ would require >100 transits using NIRSpec to observe in the NIR. However, detection of both methane (CH$_4$) and carbon dioxide (CO$_2$) may indicate a biologically-driven disequilibrium characteristic of anoxic biospheres \citep{Krissansen2018}, that may have been present on the early Earth, prior to the rise of O$_2$. 10 transits of TRAPPIST-1e observed with JWST's NIRSpec Prism may be sufficient to detect CO$_2$ and constrain the CH$_4$ abundance sufficiently well to rule out known, nonbiological CH$_4$ production scenarios \citep{Krissansen2018}. Meadows et al. (in prep) also find that the combination of CH$_4$ and CO$_2$ would be detectable for TRAPPIST-1e even if it possessed a modern Earth-like atmosphere, due to the increased photochemical lifetime of CH$_4$ expected for exoplanets in M dwarf systems (e.g. \cite{Segura2005}).  False positives for this biological CO2/CH4 disequilibrium could come from volcanism from a more reducing mantle, but this could be made less likely via a significant non-detection of CO, which may also be possible with JWST \citep{Krissansen2018}. However, the thorough survey of  environmental context needed to make a definitive claim of life's impact on a planetary environment will likely not be possible with JWST, and any tantalizing discovery of possible signs of life will require more capable future missions to verify.   


\section{The TRAPPIST-1 JWST Initiative}

As described in the previous sections, an ambitious JWST program targeting TRAPPIST-1 represents a unique scientific opportunity to probe the atmospheric and surface properties of temperate Earth-sized worlds, but it will inevitably face numerous challenges both from an observational and a theoretical point of view. Considering these challenges, the limited lifetime of JWST combined to the unfavorable position of the star in the sky for the telescope, the probably large delay between the end of the JWST mission and the launch of another space-based facility of similar or superior instrumental potential, the division of this ambitious program in several sub-programs carried out by independent teams, without any global coordination, is certainly not optimal as it would inevitably result in useless competitive efforts, delays between the publications of the results of one team and the decision by another team to complement them with other observations, missed opportunities, and eventually a non-optimal scientific return. Such an ambitious study that should require several hundreds of hours on the largest-ever space-based telescope, a telescope that could be fully operational for less than 6 years, makes it necessary to adopt a community-driven approach instead of a `competing small teams' one. It requires the support and contribution of a large interdisciplinary community gathering all the expertise and resources needed to (1) constantly optimize a well-defined sequential structure for the study, and to ensure (2) a strong coordination between all the teams involved on every aspect of its preparation and implementation, both on the observational (e.g. study of the instrumental limitations, data analysis techniques, complementary space-based and ground-based observations) and theoretical levels (e.g. model developments and comparison, retrieval techniques, inferences).  

In this context, we have set up a {\it TRAPPIST-1 JWST Community Initiative} open to any interested scientist. It aims to maximize the scientific return of the study of the TRAPPIST-1 system with JWST by: \begin{itemize}
\item Gathering all interested scientists into a large multi-disciplinary community that will work together on a single large effort, while keeping their basic scientific freedom (to lead their own proposals, to publish their own independent papers, to collaborate with the colleagues of their choice, etc.);
\item Setting-up and developing a website that will gather all relevant data and publications on the system (e.g. Spitzer photometry, high-res spectra, transit and occultation ephemerides);
\item Defining and constantly optimizing a sequential structure for the study of the system with JWST, and coordinating on every aspect of its preparation and implementation, both on the observational and theoretical levels;
\item Assessing the relevance and possibility of a major JWST Legacy program devoted to the study of TRAPPIST-1, and possibly implementing it.
\end{itemize}

The Initiative is supervised by a Board composed of scientists of complementary expertise that aims (1) to develop and provide the Initiative members with the required coordination mechanisms/tools, (2) to constantly evaluate the complementary nature of the different JWST proposals and their capacity to result into an optimal study of the system with JWST,  (3) to review globally the progress of the Working Groups (see below), (4) to manage the promotion of the Initiative, (5) to ensure that all members abide to the Initiative’s code of conduct. 

To optimize its activities, the scientific activities of the Initiative will be divided in several working groups (WGs), the current ones being: \begin{itemize}
\item {\it WG1 - Assessing GTO Observations}: the aim of this WG is to combine models from multiple groups to assess the possible scientific output of the GTO observations and the kind of complementary observations that could be proposed in Cycle 1 for planets b, d, e, f (the other planets are not part of GTOs). The science assessment will look at what models/environmental characteristics could be discriminated with the planned GTO observations and more extensive observations, including the significance of a null result.
\item {\it WG2 - Complementary Observations}: this WG aims to organize the ground- and space-based observations that would complement and optimize the JWST ones: transit timing/TTVs, RVs, stellar variability (to help constraining the atmospheric evaporation + photospheric + chromospheric models but also the models for the stellar contamination of the transit transmission spectra of the planets). It also aims to develop an online tool providing the most up-to-date transit/occultation ephemerides for all planets based on TTV analysis, and an online database grouping all TRAPPIST-1 data and publications that could be useful to prepare and analyze JWST observations.
\item {\it WG3 - Planetary Environments and Evolution}: this group will work on anticipated/predicted system-wide trends in atmospheric characteristics to help justify observing the entire sequence of 7 planets, rather than just one or two. This is envisioned to be an interdisciplinary group focused on atmospheres, but taking into account the modification of atmospheres via atmospheric loss as well as replenishment by outgassing. Specific questions to be addressed include: What can geophysics tell us about the possible atmospheric compositions of the planets, considering different bulk compositions and outgassing scenarios? One of its output will be a table listing possible atmospheric scenarios for the planets and the observations required to falsify these hypotheses, eventually resulting into a global and sequential plan for the atmospheric characterization of the planets.
\item {\it WG4 - JWST Synthetic Data Generation and Analysis}: working group interested in synthetic JWST data generation and analysis to support observation planning.
\end{itemize}
This structure is by essence flexible, and it will adapt to the evolution of the global study of TRAPPIST-1. Each WG will work by teleconferences and internet exchanges (emails, Slack, etc). In  the long term, some workshops could also be organised to optimize the efficiency of the WGs.  

The main tools put at the disposal of the Initiative members is a website (\url{https://nexss.info/community/trappist-1}) where interested scientists can sign up to join the Initiative and to choose the WG(s) to which they would like to contribute. This website will include contact information for all Initiative members, for the Board and for the WGs; 
an online transit/occultation ephemerids tool that will be regularly updated based on the latest TTV results; and access to an online data and publication database relative to TRAPPIST-1.
In addition to the website, the Initiative will provide to its members private Slack channels where members will be able to post their proposals and publication intentions, their proposals, a draft of their publications, communicate within a WG, etc.  

Any comments and suggestions on the structure and activities of the Initiative are welcome. They can be addressed to the board via the email \href{mailto:t1jwstci\textunderscore board@u.washington.edu}{t1jwstci\textunderscore board@u.washington.edu}.

\section*{Acknowledgments}
This work has made use of data from the European Space Agency (ESA) mission
{\it Gaia} (\url{https://www.cosmos.esa.int/gaia}), processed by the {\it Gaia}
Data Processing and Analysis Consortium (DPAC,
\url{https://www.cosmos.esa.int/web/gaia/dpac/consortium}). Funding for the DPAC
has been provided by national institutions, in particular the institutions
participating in the {\it Gaia} Multilateral Agreement.
This publication makes use of data products from the Two Micron All Sky Survey, which is a joint project of the University of Massachusetts and the Infrared Processing and Analysis Center/California Institute of Technology, funded by the National Aeronautics and Space Administration and the National Science Foundation.

\printbibliography[title={Bibliography}] 

\newpage
\appendix

\section{The 40pc M+L dwarfs sample} \label{annex1}
To develop our catalog of M- and L-dwarfs within 40pc, we started with the 35,781 objects in the Gaia DR2 catalog \citep{Gaia2016, Gaia2018} with a trigonometric parallax $d \ge 25$ mas. 
For each of them, we (1) corrected the parallax from the small systematic offset of -82 $\mu$as identified by \textcite{Stassun2018}, and added quadratically the error on this offset (33 $\mu$as) to the Gaia parallax error;
(2) computed the J2000 equatorial coordinates considering only the proper motion as measured by Gaia (the epoch of Gaia DR2 coordinates is J2015.5); 
(3) computed the absolute magnitude $M_G$ from the apparent $G$-band magnitude and the distance modulus measured by Gaia; 
(4) computed an estimate of the effective temperature $T_{eff}$ based on the empirical law $T_{eff}(M_G)$ of \textcite{PM2013} (hereafter PM2013) and assuming a systematic error of 150K added quadratically to the error propagated from the error on $M_G$. 
We then discarded all objects with $M_G < 6.5$ or a color $G_{BP} - G_{RP} < 1.5$ to keep only dwarf stars later than $\sim$K9-type. We also discarded objects missing a $G_{BP} - G_{RP}$ color in Gaia DR2. We ended up with 21,137 potential nearby M- and L-dwarfs.

We then cross-matched each of these objects with the 2MASS point sources \citep{2MASS} within 2'/$d$ (=3'' at 40pc). 
This 1/$d$ dependency of the search radius aims to take into account that, for the nearest stars, the astrometric position at a given time can differ significantly from the one computed  by  correcting the J2015.5 position from the proper motion because of the significant  amplitude of their 3D motion, meaning that their radial velocity should also be considered. 
For instance, there is an astrometric difference $>$30''  between the J2000 position of Proxima Centauri as measured by 2MASS and the one computed from its Gaia DR2 J2015.5 coordinates and proper motion.

For each 2MASS object falling within 2'/$d$ of a selected Gaia DR2 object, we computed two estimates of $T_{eff}$, one based on the $T_{eff}(G-H)$ empirical relationship of PM2013 assuming a systematic error of 150K, and one based on the $T_{eff}(M_H)$  empirical relationship of \textcite{F2015} (hereafter F2015), assuming a systematic error of 100K, and with the absolute magnitude $M_H$ computed from the apparent $H$-band magnitude measured by 2MASS and from the distance modulus measured by Gaia. We then  computed for each Gaia - 2MASS couple the following metric: \\
\begin{equation}
\begin{array}{l}
\left ( \frac{T_{eff}(M_G)-<T_{eff}>}{\sigma_{T_{eff}(M_G)}} \right )^2 + 
\left ( \frac{T_{eff}(G-H)-<T_{eff}>}{\sigma_{T_{eff}(G-H)}} \right )^2 + \\
\left ( \frac{T_{eff}(M_H)-<T_{eff}>}{\sigma_{T_{eff}(M_H)}} \right )^2 + 
\left (  \frac{\delta_{position}}{\sigma_{\delta_{position}}}\right )^2
\end{array}
\end{equation} where $<T_{eff}>$ is the mean of the three temperature estimates and where
\begin{eqnarray} 
\delta_{position} &=& \arccos{(\sin{\delta_1}\sin{\delta_2} } \\
&+& \cos{\delta_1}\cos{\delta_2}\cos{(\alpha_1 - \alpha_2)}), 
\end{eqnarray} $\alpha_i$ and $\delta_i$ being the right 
ascension and declination of the star $i$, and where $\sigma_{\delta_{position}}$ is the error on the position difference between the two objects computed from propagation of the errors on $\alpha$ and $\delta$ quadratically summed to an error of 40''/$d$ (one third of the search radius) to take into account the significant 3D motion of the nearest stars.

Each Gaia DR2 object was cross-matched with the 2MASS object within 2'/$d$ that minimized its metric function, i.e. with the nearest position AND the best match in terms of $T_{eff}$ as derived from $M_{G}$, $M_{H}$ and $G-H$. 
For 3,654 objects, no cross-match was found, i.e. no 2MASS object was found within 2'/$d$. 
For the remaining 21,137 - 3,654 = 17,483 objects, the three temperature estimates $T_{eff}(M_G)$, $T_{eff}(G-H)$, and $T_{eff}(M_H)$ were compared between each other. 
In case of discrepancy at more than $2\sigma$, the object was discarded. 2,901 objects were rejected through this temperatures comparison, leaving 14,582 objects.

For these objects, we (1) derived an estimate of the spectral type $SpT$ by inverting the empirical relationship $T_{eff}(M_H)$ of F2015, assuming an internal error of 113K for it; 
(2) computed an estimate of the $I_c$-band magnitude from the 2MASS $J$-band magnitude and the spectral type  estimate using online tables with empirical colors as a function of spectral type\footnote{http://www.stsci.edu/~inr/intrins.html}; 
(3) computed the photometric precision that should be achieved by TESS within 30min basing on the apparent  $I_c$ magnitude\footnote{https://tasoc.dk/docs/release\textunderscore notes/tess\textunderscore sector\textunderscore04\textunderscore drn05\textunderscore v03.pdf}; 
(4) used our Exposure Time Calculator of SPECULOOS to compute the typical photometric precision that should be achieved by a SPECULOOS telescope as a function of the apparent $J$-band magnitude and the spectral type;  
(5) computed a $J$-band bolometric correction $BC_J$. For $SpT$ later than M6.6, we used the $BC_J(SpT)$ relationship of F2015  (assumed internal error = 0.163), and for earlier stars we used the $BC_J(T_{eff})$ relationship of PM2013 assuming an internal error of 0.2; 
(6) computed an estimate of the bolometric luminosity $L_{bol}$ (+ error) from $M_J$ and $BC_J$; 
(7) computed an estimate of the radius $R\ast$ (+ error) from $L_{bol}$, $T_{eff}$, and the relationship $L_{bol} = 4 \pi R_\ast^2 T_{eff}^4$;
(7) derived an estimate of the mass $M_\ast$ (+ error) from the empirical relationship of \textcite{Mann2019} (assumed internal error = 0.03\%) for objects earlier than L2.5, assuming them to be low-mass stars \citep{Dieterich2014}. 
For later objects (i.e. brown dwarfs), we assumed the following relationship to derive a crude estimate of the mass: $M_\ast = 0.075 - (SpT-12)*0.0005 M_\odot$, with an assumed error of 20\%. 
As the spectral type of a brown dwarf is not only correlated with its mass but also its age, this relationship has no ambition to be accurate at all. 
It just aims to represent that, statistically speaking, a hotter brown dwarf tends to be more massive than a colder one. 

At that stage, we rejected another batch of 478 objects for which at least one of the following conditions was met: \begin{itemize}
    \item The computed radius was smaller than 0.07 $R_\odot$, i.e. too small for an ultracool dwarf \citep{Dieterich2014}.
    \item The number of 2MASS objects within 2'' was over 250 and the $K$ magnitude larger than 12.5, making likely a confusion case (galactic disk + bulge).
    \item The inferred spectral type  was later than M5.5 [L0], $K$ was larger than 8 (so no saturation in 2MASS images), and still the $J - K$ color was smaller than 0.6 [1.0], suggesting a wrong cross-match or a confusion case.
    \item The inferred mass was smaller than 0.2 $M_\odot$ while the inferred radius was larger than 0.4 $R_\odot$, i.e. too large for a low-mass M-dwarf.
\end{itemize}
We ended up eventually with a 40pc M+L dwarfs catalog containing 14,104 objects. 
Still, we noticed that some well-known nearby late-M and L-dwarfs were missing from the catalog because they had no parallax in Gaia DR2, including very nearby objects like Scholtz star (M9.5+T5.5 at 6.7pc, \citealt{Burgasser2015}), Luhman-16 (L7.5+T0.5 at 2.0pc, \citealt{Luhman2013}), or Wolf 359 (M6.0 at 2.4pc, \citealt{Henry2004}). 
We thus cross-matched our catalog with the spectroscopically verified sample of M6-L5 ultracool dwarfs (UCDs) within 25 pc published recently by \textcite{Gagliuffi2019}. 
For each object within this sample and not present in our catalog, we (1) used  the $T_{eff}(SpT)$ empirical relationship of F2015 to estimate the effective temperature; 
(2) used the same procedure than for the Gaia DR2 objects to estimate the bolometric correction, the luminosity, the radius, the mass, and the SPECULOOS and TESS photometric precisions. 
We discarded objects flagged as close binary in the catalog of \textcite{Gagliuffi2019}, 
objects with an inferred size smaller than 0.07 $R_\odot$, objects with an inferred mass greater than 0.125 $M_\odot$ (too massive for an ultracool dwarf, suggesting a blend or a close binary case), and objects later than M9.0 with a $J-K$ color index smaller than 1.0 and a $K$ magnitude larger than 8. 
At the end, this procedure added 59 objects to our catalog, for a total of 14,163. 
As our initial goal was to build the target list of SPECULOOS that is, we did not try to recover stars earlier than M6 absent from our catalog because they do not have a Gaia DR2 parallax. 
Extrapolating our results for ultracool dwarfs to earlier M-dwarfs, we estimate that there must be at most a few hundreds of them in the case. 

\begin{figure*}
    \centering
    \includegraphics[width =\columnwidth]{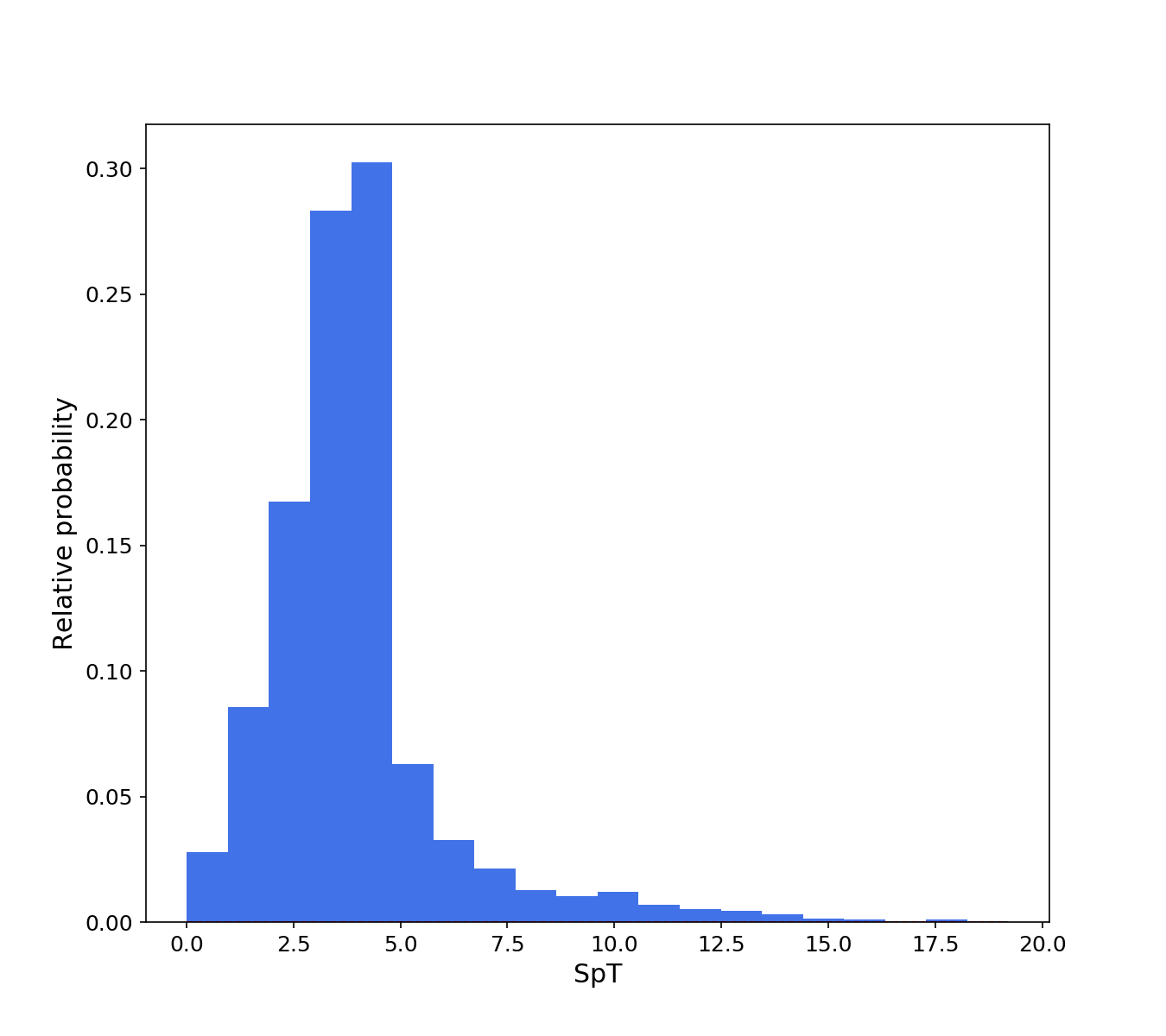}
    \includegraphics[width =\columnwidth]{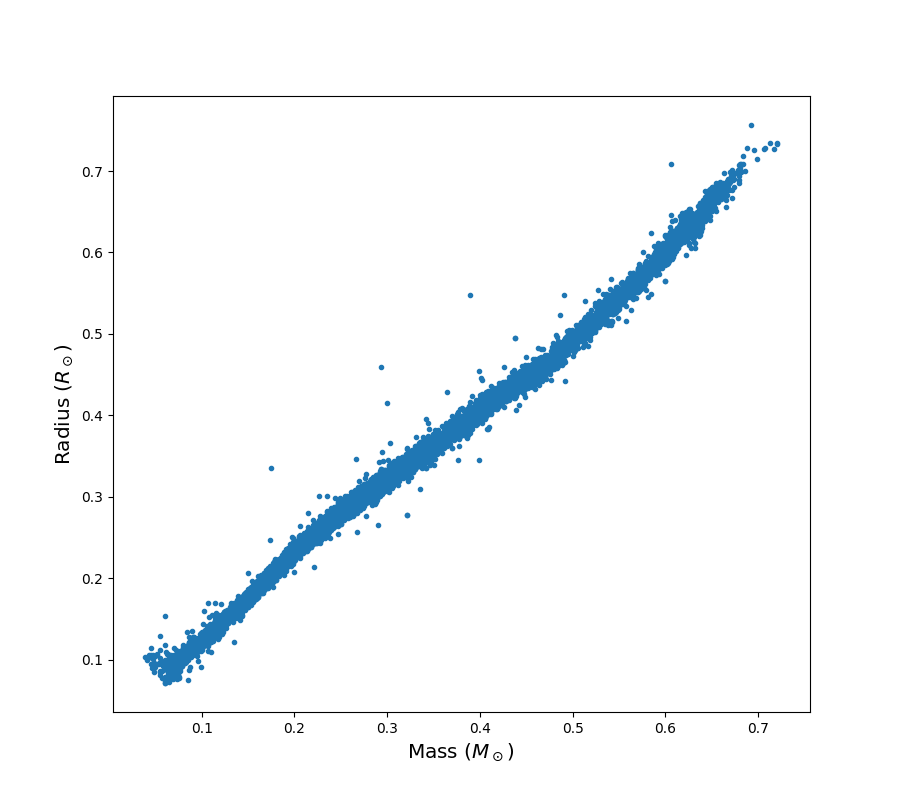}
    \caption{Spectral type distribution ($left$) and mass-radius diagram ($right$) for our 40pc ML-dwarfs catalog. }
    \label{fig:cat}
\end{figure*}

Figure \ref{fig:cat} shows the spectral type distribution and the mass-radius diagram of the 14,163 objects. One can notice that our catalog of nearby M-dwarfs is dominated by $\sim$M4-type objects, in agreement with earlier results (e.g. \citealt{Henry2018}). 

We then computed a metric for each target of our catalog  to quantify its value as target of an ambitious 200hr JWST/NIRSPEC program aiming to study the atmospheric composition of a putative transiting `Earth-like' (same mass, same size, same atmospheric composition with a mean molecular mass of 29amu, same irradiation from the host star, same Bond albedo of 0.3) planet. 
For the typical amplitude of the transmission signals, we used the following formula \citep{Winn2010}:
\begin{equation}
\Delta \delta = 2 N_H \delta \big( \frac{H}{R_\oplus} \big)
\end{equation} where $H$ is the atmospheric scale height assuming an isothermal atmosphere of temperature equal to the equilibrium temperature of the planet, $\delta$ is the transit depth, and $R_\oplus$ is the Earth's radius. 
We assumed a value of 5  for $N_H$, the number of scale heights corresponding to a strong molecular transition. 
For the assumed planets, the orbital distance corresponding to an Earth-like irradiation was computed basing on the stellar luminosity, the corresponding orbital period was computed using Kepler's third law combined to the stellar mass estimate, and the duration of a central transit was computed using Formula 15 from \cite{Winn2010}. 
We then computed the JWST/NIRSPEC  (Prism mode) noise at 2.2 $\mu$m for a spectral bin of 100nm and for an exposure sequence with the same duration than the transit using the online tool PandExo \citep{Batalha2017}. 
We assumed a floor noise per transit of 30ppm that we added quadratically to the white noise estimate of PandExo. 
We computed the number of transits observed within the 200hr JWST program, assuming for each transit observation a duration equal to the transit duration plus 2.5hr (for pointing, acquisition, plus out-of-transit observation). 
The noise per transit was then divided by the square root of the number of observed transits, and we added quadratically to the result an absolute floor noise of 10ppm. 
At the end, we obtained for each target a transmission signal-to-noise (SNR) by dividing the transmission amplitude by the global noise. Figure \ref{fig:jwst} (left) shows the resulting $SpT-SNR$ distribution. 
In this Figure, we draw a line at $SNR$ = 5, assuming this value to be an absolute minimum for deriving meaningful constraints on the atmospheric composition of our assumed Earth-like' planets. 
TRAPPIST-1 is shown as a red star symbol in the Figure. There are 101 objects -including TRAPPIST-1- that have a SNR $\ge$ 5, and only 44 that have a SNR larger than TRAPPIST-1.  

We performed the same exercise for occultation spectrophotometry with JWST/MIRI (imaging mode) in the F1000W filter ($\lambda = 10\mu$m, $\Delta\lambda = 2 \mu$m), assuming a 25hr program aiming to detect the thermal emission of an atmosphere-less Earth-sized planet with an irradiation four times larger than Earth, as TRAPPIST-1b. 
Black body spectra were assumed for the star (with the $T_{eff}$ estimated the spectral type, as describe above) and planet (with $T_{eff}$ = 490K = equilibrium temperature for $A_B$ = 0.1 and an inefficient heat distribution from the day- to the night-side). We derived a $K$-$W3$(SpT) relationship  from the data of \textcite{Schmidt2015} and literature $K$-  and $W3$-magnitude measurements for nearby M-  and L-dwarfs. 
We used the online JWST Exposure Time Calculator\footnote{https://jwst.etc.stsci.edu} to estimate the photometric precision of MIRI to be $\sim$100ppm for TRAPPIST-1 ($W3 = 9.6$) in the F1000W filter for an exposure time of 30min, and rescaled this error for each target based on its $W3$-magnitude and the duration of a central occultation. Here too, we assumed floor noises of 30ppm per transit and 10ppm for the whole program.  
286 objects have a $SNR \ge 5$. We obtained a $SNR \sim 6.6$ for TRAPPIST-1, and larger values for 131 objects of our 40pc M+L dwarfs catalog, most of them being of later-type (Figure \ref{fig:jwst}, right). 

\begin{figure*}
    \centering
    \includegraphics[width =\columnwidth]{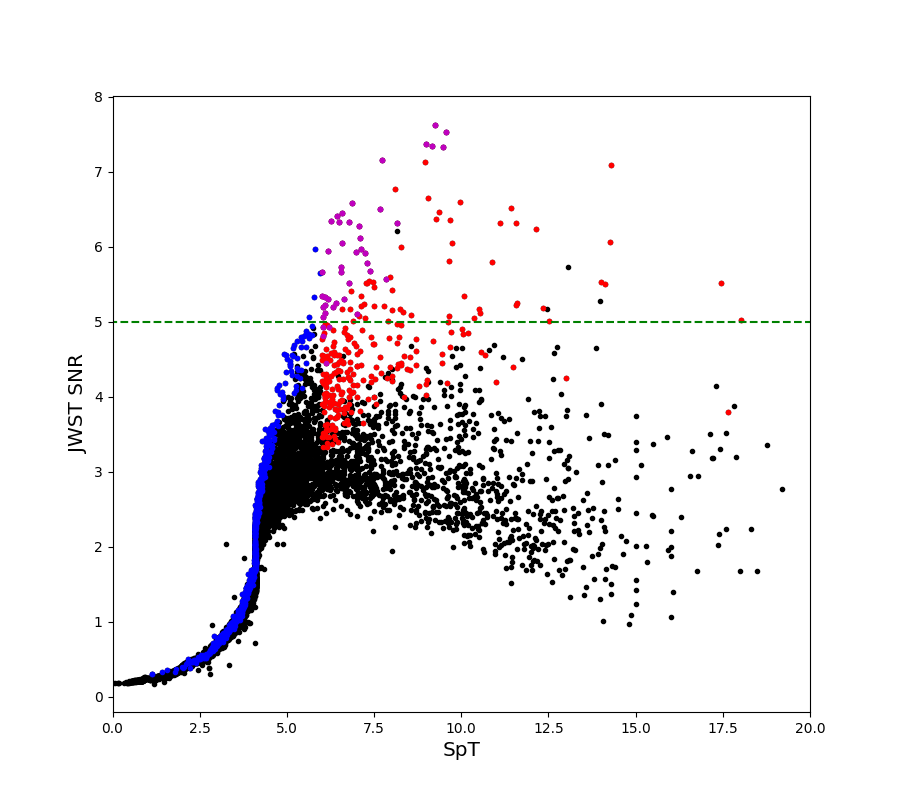}
    \includegraphics[width =\columnwidth]{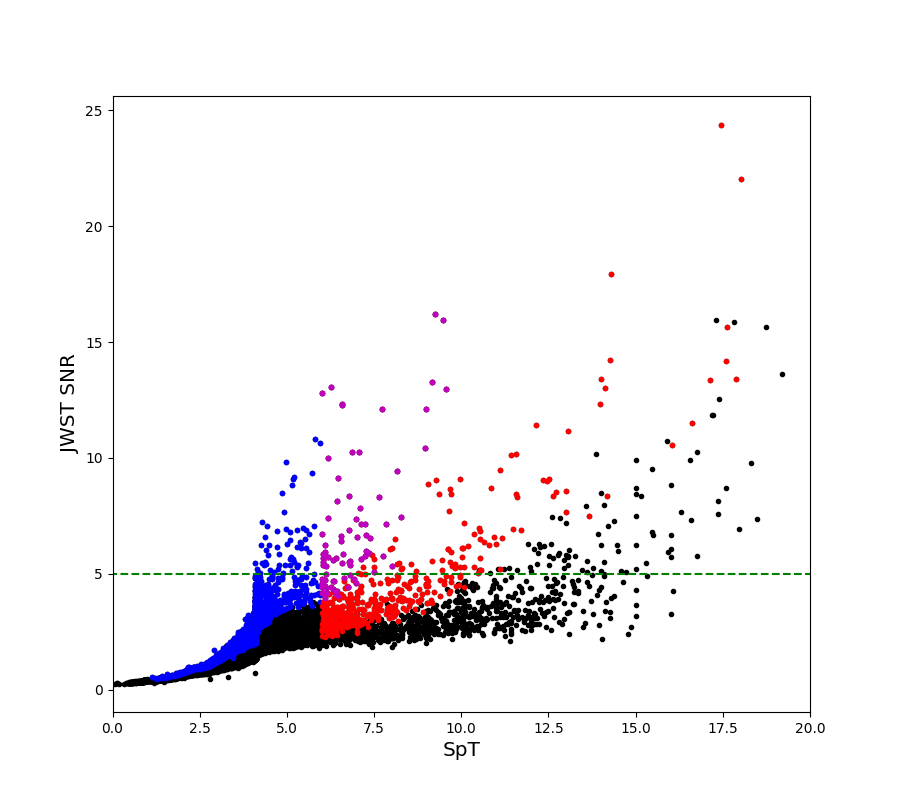}
    \caption{Same as Fig. \ref{fig:jwst} but with
    targets for which a transit detection would be within reach of TESS, SPECULOOS,
    or both are overplotted as blue, red, and magenta symbols, respectively.}
    \label{fig:jwst_det}
\end{figure*}

For each target, we then assumed two 27d-long photometric monitoring by TESS, and we computed the SNR for the detection of Earth-sized planets with irradiations of one and four times the Earth's,  considering our computed TESS photometric precision estimate for the star combined to the mean number of observed transits,  and their depths and durations (assuming a central transit). 
A floor noise of 50ppm was assumed per transit, and none for the phase-folded photometry. 
If the computed SNR was $\ge$8, a detection within the reach of TESS was inferred. 
We applied the same procedure for SPECULOOS, considering only targets with spectral type later than M5.5, assuming this time a 100hr-long photometric monitoring and a floor noise of 500ppm per transit,  a realistic floor noise for high-precision ground-based photometry from an good astronomical site and a state-of-the-art equipment. 
Here too, a detection was deemed possible when the computed SNR was $\ge$8. Figure \ref{fig:jwst_det} summarizes our results. 
It is similar to Fig. \ref{fig:jwst} except that potential `Earth-like' planets whose detection is within reach of TESS and SPECULOOS are overplotted. 
Among the 101 targets that are well-suited for transmission spectroscopy of an `Earth-like' planet, TESS and SPECULOOS are able to detect the planet for 43 and 96 objects, respectively. This larger potential of SPECULOOS is due to the fact that transmission spectroscopy of  potentially habitable Earth-sized planets is possible with JWST only for stars that are both very nearby ($< \sim 15$ pc) and very small ($< \sim 0.15 R_\odot$), as TESS detection potential falls down at spectral type $\sim M5$ \textcite{Sullivan2015} unlike the one of SPECULOOS that was specifically designed to detect Earth-sized planets transiting ultracool dwarf stars \citep{Gillon2013, Burdanov2018, Delrez2018b, Gillon2018}.

\end{document}